\documentclass[aps,prb,twocolumn,superscriptaddress,floatfix,citesort,showpacs]{revtex4}
\usepackage{epsfig}

\begin{document}

\title{Dynamical heterogeneities close to a colloidal gel}
\author{Antonio M. Puertas}
\affiliation{Group of Complex Fluids Physics, Department of Applied Physics, University of Almeria, 04120 Almeria, Spain}
\author{Matthias Fuchs}
\affiliation{University of Konstanz, D-78457 Konstanz, Germany}
\author{Michael E. Cates}
\affiliation{Department of Physics and Astronomy, The University of
Edinburgh, EH9 3JZ, UK}
\date{\today}

\begin{abstract}

Dynamical heterogeneities in a colloidal fluid close to gelation are studied by means of computer simulations. A clear distinction between some fast particles and the rest, slow ones, is observed, yielding a picture of the gel composed by two populations with different mobilities. Analyzing the statics and dynamics of both sets of particles, it is shown that the slow particles form a network of stuck particles, whereas the fast ones are able to move over long distances. Correlation functions show that the environment of the fast particles relaxes much faster than that of the slow ones, but at short times the bonds between fast particles are longer lived due to the flexibility of their structure. No string-like motion is observed for the fast particles, but they occupy preferential sites in the surface of the structure formed by the slow ones.

\end{abstract}

\maketitle

\newpage

\section{Introduction}

Colloidal gelation occurs in a system with short range attractive interactions, at high interaction strength for a wide range of density. Upon increasing the attraction, the system falls out of equilibrium, forming a network of particles with voids, that spans through the system. As observed by light scattering, a static low angle peak becomes apparent, signaling the formation of a ramified structure \cite{russel89,haw95}. This process has no analogue in atomic systems and presents a challenge in condensed matter physics. It is, thus, receiving increasing attention over the past few years \cite{imhof95,fabbian99,bergenholtz99,bergenholtz99b,dawson01,segre01,pham02,dawson02,eckert02,sciortino02,bergenholtz03,sedgwick03,pham04}. Its application in industrial processes and other sciences makes gelation even more intriguing \cite{russel89}. Also, understanding gelation at low densities may lead to a full description of colloidal aggregation \cite{kroy03,weitz84,weitz85,lin90,carpineti92}.

Colloidal gelation has been tackled recently by means of mode coupling theory (MCT) \cite{gotze91,gotze92}, which addresses the glass transition both in atomic and colloidal fluids \cite{fabbian99,bergenholtz99,bergenholtz99b,dawson01}. Within MCT, gelation reduces to a regular ergodic to non-ergodic transition, similar to the glass transition taking place at high density \cite{megen94,bartsch97,megen98,beck99}. As predicted by MCT, the gel transition occurs for all densities, for higher attraction strength the lower the density. Using molecular dynamics simulations, the predictions from MCT have been tested at high density, far from the percolation line, and good agreement has been found, showing that MCT correctly describes this transition \cite{puertas02,foffi02,zaccarelli02,puertas03}. However, at low density this description is still under debate \cite{kroy03,sedgwick03,zaccarelli01,zaccarelli03}.

Computer simulations of the glass transition at high density have shown also good agreement with the predictions from MCT \cite{kob95a,kob95b,nauroth97}. However, some differences have also been pointed out, especially the so-called dynamical heterogeneities \cite{kob97,donati99a,donati99b,doliwa00}. Particles with different mobilities are detected in the glass and in the fluid close to the glass, and string-like motions of particles have been reported. These heterogeneities in the system, which are more pronounced as the glass transition is approached, have been observed also experimentally \cite{weeks00,weeks02}. In this work, by means of computer simulations, we study if dynamical heterogeneities are also present in fluid states close to gelation. The system under study is a colloid-polymer mixture, modeled by the Asakura-Oosawa interaction potential between the colloids, which has been shown previously to undergo a non-ergodic transition at high attraction strength \cite{pham02,bergenholtz03,pham04,puertas02,puertas03}. 

Analyzing the distribution of squared displacements, we show that dynamical heterogeneities indeed exist in the gel, and, in a simplified view, two sets of particles with different mobilities can be recognized. The fast particles have fewer neighbors than the slow ones, and are also smaller (a polydisperse system is simulated to prevent crystallization). Since the exchange of particles between the fast and slow populations is very slow, we describe the system as composed by two distinct populations of particles, with different mobilities. The slow particles form a percolating network with the same structure as the overall system, very stable in time, whereas the structure formed by the fast ones relaxes faster. In the surface of the network formed by the slow particles, there exist preferential sites for the fast ones, which are occupied when vacant. The environment of the particles shows that fast particles are surrounded by other fast particles mainly, whereas slow ones bond to other slow ones. The stiffness of the structure of the slow particles makes it more fragile, and, at short times, bonds between short particles are shorter lived than those between fast ones.

In this paper we analyze solely the dynamic heterogeneities, and refer the reader to a previous paper where the MCT predictions are tested for this particular system \cite{puertas02,puertas03}. In section II we present the details of the system and of the simulation method. Section III deals with the results; first we present the distribution of squared displacements, where the dynamic heterogeneity is clearly inferred. Then, the static properties of both populations are analyzed and their dynamics are studied. We then discuss the results and present the relevant conclusions of this work. 

\section{Simulation details}

Molecular dynamics simulations were performed to analyze the dynamics of single particles in the system close to gelling. The system is comprised of 1000 soft core polydisperse particles, with a short range attraction given by the Asakura Oosawa potential, modeling a mixture of colloids with non-adsorbing polymer. The core-core repulsion is given by:

\begin{equation}
V_{sc}(r)\:=\:k_BT \left(\frac{r}{a_{12}}\right)^{-36}
\end{equation}

\noindent where $a_{12}=(a_1+a_2)/2$, with $a_1$ and $a_2$ the radii of the interacting particles. Particle sizes are distributed according to a flat distribution of half-width $\delta=0.1 a$, where $a$ is the mean radius; $a_i \in [0.9a,1.1a]$. The attraction induced by the polymers, extended to take polydispersity into account, reads \cite{asakura54,dijkstra99,mendez00}:

\[ V_{AO}(r) \:=\: -k_BT \phi_p \left\{\left[\left(\bar{\eta}+1\right)^3 -\frac{3r}{4\xi} \left(\bar{\eta}+1\right)^2+\frac{r^3}{16\xi^3}\right]+
\right.\]
\begin{equation}\label{pot} 
\left.+\frac{3\xi}{4r} \left(\eta_1-\eta_2\right)^2 \left[\left(\bar{\eta}+1\right) -\frac{r}{2\xi} \right]^2\right\}
\end{equation}

\noindent for $2(a_{12}+\xi/5) \leq r \leq 2(a_{12}+\xi)$ and 0 for larger distances. Here, $\eta_i=a_i/\xi$; $\bar{\eta}=(\eta_1+\eta_2)/2$, and $\phi_p$ is the volume fraction of the polymer. Note that the range of the potential is given by $\xi$, the polymer size, and its strength is proportional to $\phi_p$. For $r \leq 2(a_{12}+\xi/5)$, the potential is parabolic, connected analytically to $V_{AO}$ at $r= 2(a_{12}+\xi/5)$, with the minimum at $r=2 a_{12}$. The total potential, thus, has a minimum very close to $r=2 a_{12}$.

Since liquid-gas separation would prevent us from accessing a huge area of the density-attraction strength plane, a long-range barrier has been added to the interaction potential. This barrier has the form:

\begin{equation}
V_{bar}(r)\:=\:k_BT\left\{\left(\frac{r-r_1}{r_0-r_1}\right)^4-2\left(
\frac{r-r_1}{r_0-r_1}\right)^2+1\right\}
\end{equation}

\noindent for $r_0\leq r \leq r_1$ and zero otherwise. The limits of the barrier were set to $r_0=2(a_{12}+\xi)$, and $r_1=4a$, and its height is only $1 k_BT$\footnote{This energy is equal to the attraction strength at $\phi_p=0.0625$.}. The barrier raises the energy of a dense phase, so that liquid gas separation does not take place. 

\begin{figure}
\psfig{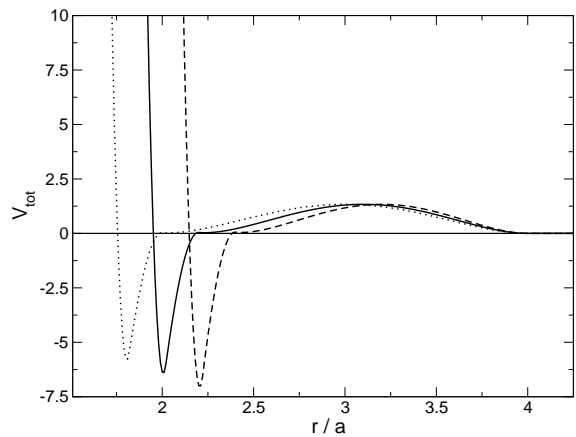}
\caption {\label{potential}
Total pair interaction potential $V_{tot}$ as function of the radial distance $r=|{\bf r}_1-{\bf r}_2|$ for three different particle pairs; a pair of particles with minimal radii $a_1=a_2=a-\delta$, one with average radii $a_1=a_2=a$, and one with maximal  $a_1=a_2=a+\delta$ (from left to right).}
\end{figure}

The resulting total interaction potential, $V_{tot}=V_{sc}+V_{AO}+V_{bar}$, which is analytical everywhere, is shown in Fig. \ref{potential}. In order to show the effect of polydispersity, the extreme cases, as well as the average one are presented. For small particles, not only the core interaction is altered, but also the attraction is weaker, whereas big particles interact with a stronger attraction. The maximal difference between interacting pairs of particles is $1.25\,k_BT$. 

In our simulations, lengths are measured in units of the average radius, $a$, and time in units of $\sqrt{4a^2/3v^2}$, where the thermal velocity $v$ was set to $\sqrt{4/3}$. Equations of motion were integrated using the velocity-Verlet algorithm, in the canonical ensemble (constant NTV), to mimic the colloidal dynamics, with a time step equal to $0.0025$. Every $n_t$ time steps, the velocity of the particles was re-scaled to assure constant temperature. No effect of $n_t$ was observed for well equilibrated samples. Equilibration of the samples was tested by the trends of the energy of the system and other order parameters \cite{puertas03}, and by the independency of the correlation functions on the initial time. For the highest polymer fraction studied here, $\phi_p=0.42$, equilibration ran for $5\cdot 10^4$ time units, corresponding to $2\cdot 10^7$ time steps.

The range of the attraction is set to $2\xi=0.2$. The density of colloids is reported as volume fraction, $\phi_c=\frac{4}{3}\pi a^3 \left(1+\left(\frac{\delta}{a}\right)^2\right) n_c$, with $n_c$ the colloid number density, and the attraction strength is given by the polymer volume fraction, $\phi_p$. 

\section{Results and Discussion}

Dynamical heterogeneities in a system can be detected analyzing the distribution of the squared displacement of all particles at a fixed time. For homogeneous colloidal fluids this distribution is single peaked, the width depending on the self diffusion coefficient of the particles and on time \cite{hansen86}. However, this is not the case when populations of particles with different mobilities are present. 

\begin{figure}
\psfig{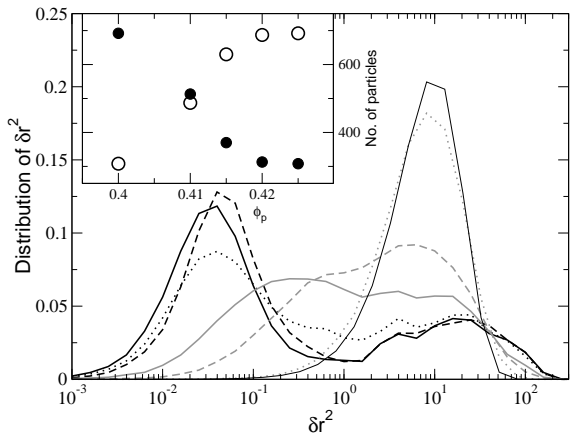}
\caption {\label{fig1} Distribution of squared displacement for $\phi_p=0.30$ (gray dotted line), $\phi_p=0.40$ (gray dashed line), $\phi_p=0.41$ (gray solid line), $\phi_p=0.415$ (black dotted line), $\phi_p=0.42$ (black dashed line), and $\phi_p=0.425$ (black solid line). In all cases $\langle \delta r^2\rangle=10 a^2$. The thin black line is the theoretical distribution for a system of non-interacting Brownian particles. Inset: Number of slow particles (open circles) and fast particles (closed circles) in the system as a function of the polymer fraction. Gelation is estimated to occur at $\phi_p=0.4265$.}
\end{figure}

In Fig. \ref{fig1} the distribution of squared displacements\footnote{In order to tackle small and big displacement, a logarithmic binning was used to calculate the distribution.} is plotted for systems at different states for increasing attraction strength, from time zero to time $t^*$ defined by $\langle \delta r^2(t^*)\rangle=10 a^2$. At low polymer fraction, the distribution is very similar to that of a system composed of non-interacting Brownian particles \cite{dhont}. However, as the polymer concentration, $\phi_p$, is increased the distribution becomes broader and two peaks are apparent, one caused by 'slow particles' at $\delta r^2 \approx 5 \cdot 10^{-2} a^2$, of the order of the interaction range, $2\xi=0.2$, and another one at $\delta r^2 \approx 20 a^2$ signaling a population of 'fast particles'. Note that the mean squared displacement is $\langle \delta r^2\rangle=10 a^2$ in all cases. The values of $t^*$ are given in Table I for the states presented in Fig. \ref{fig1}.

\vspace{0.5cm}
\begin{center}
\begin{tabular}{ccc} \hline\hline
 $\hspace{0.25cm}$ $\phi_p$ $\hspace{0.25cm}$ & $\hspace{0.25cm}$ $t^*$ $\hspace{0.25cm}$ & $\hspace{0.25cm}$ $\Delta t$ $\hspace{0.25cm}$ \\ \hline
 0.30 & 29.26 & 2.55 \\
 0.40 & 981.6 & 70.42 \\
 0.41 & 1599 & 120.44 \\
 0.415 & 2162 & 132.78 \\
 0.42 & 2871 & 196.18 \\
 0.425 & 4911 & 289.85 \\ \hline\hline
\end{tabular}
\end{center}
{\small Table I: Time $t^*$, when the mean squared displacement is equal to $10a^2$, and interval between consecutive correlators, $\Delta t$.}
\vspace{0.5cm}

The minimum in the distribution at $\delta r^2= a^2$ permits us to establish an unambiguous way of distinguishing between 'fast' and 'slow' particles, instead of using an arbitrary number of fastest or slowest particles, as done in the study of glasses \cite{donati99b}. The number of slow and fast particles is plotted in the inset to the figure as a function of the polymer fraction. Indeed, as the attraction strengthens more slow particles and fewer fast ones are present in the system. The gel transition was estimated to take place at $\phi_p=0.4265$ by the power-law divergence of the time scale \cite{puertas03}. It is interesting to note that extrapolation of this plot yields a noticeable fraction of fast particles even in the gel.

The structure and dynamics of both types of particles, fast and slow, are analyzed in the following subsections. First, we will study the static properties, such as the distribution of fast and slow particles, and will try to correlate the character with other properties of the particle. Then, we will pay attention to the motion of single particles as well as its environment and the relaxation of the structures formed by the particles. Finally, we will discuss the results, comparing them with other observations and theory.

\subsection{Static properties}

\begin{figure}
\psfig{file=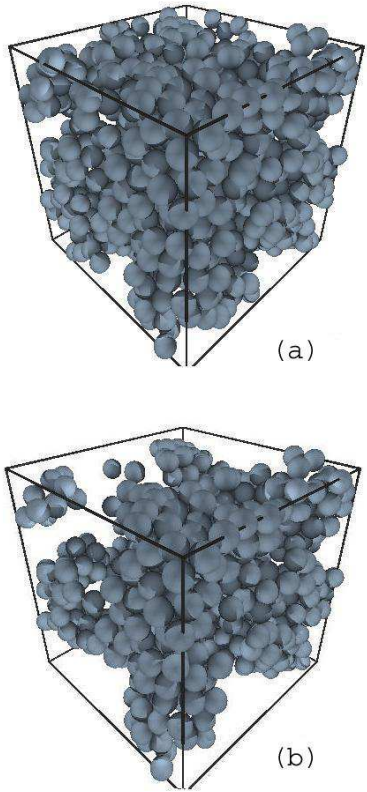,width=7.5cm}
\caption {\label{gel} Image of the system for $\phi_c=0.40$ and $\phi_p=0.42$ for the whole system (upper panel). Only the slow particles are shown in the lower panel.}
\end{figure}

The system at $\phi_c=0.40$ and $\phi_p=0.42$, which has a strong distinction between fast and slow particles, is a heterogeneous network of particles, with voids and tunnels, as presented in Fig. \ref{gel}a. The distribution of the slow particles in the system is studied in the lower part of the figure, where only these particles have been shown. A percolating network of particles is again formed, with bigger voids than in the complete system. However, since the structure of the system is so intricate, these three dimensional presentations are unclear. 

In Fig. \ref{slices} we present four slices of width $2a$ of the system, where the fast and slow particles are differently colored, dark grey particles are fast ones and light grey are slow ones. It can be seen that the slow particles form compact structures which are connected (in the $z$ direction). The fast particles, on the other hand, are also clustered but lonely fast particles are also observed, bonded to slow ones. The overall appearance is that no clear segregation between fast and slow particles is observed.

\begin{widetext}
\begin{center}
\begin{figure}
\psfig{file=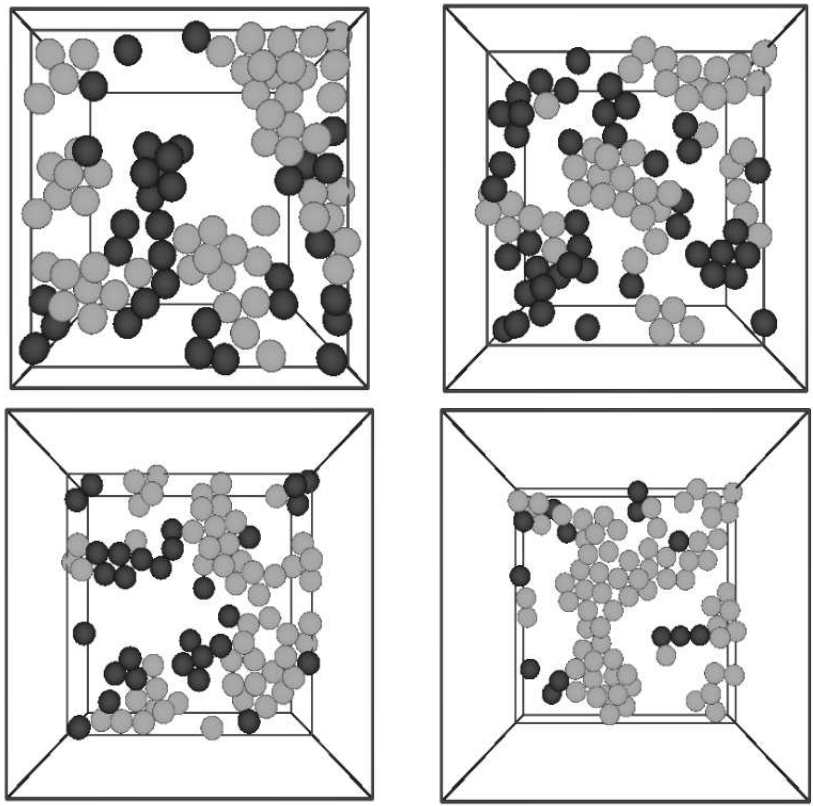,width=15cm}
\caption {\label{slices} Slices of the system at $\phi_c=0.40$ and $\phi=0.42$ for $z=7.5 a$ (upper left), $z=2.5 a$ (upper right), $z=-2.5 a$ (lower left) and $z=-7.5 a$ (lower right). Dark grey particles are fast ones ($\delta r(t^*)>  a^2$), and light grey means slow particles ($\delta r(t^*)< a^2$).}
\end{figure}
\end{center}
\end{widetext}

\begin{figure}
\psfig{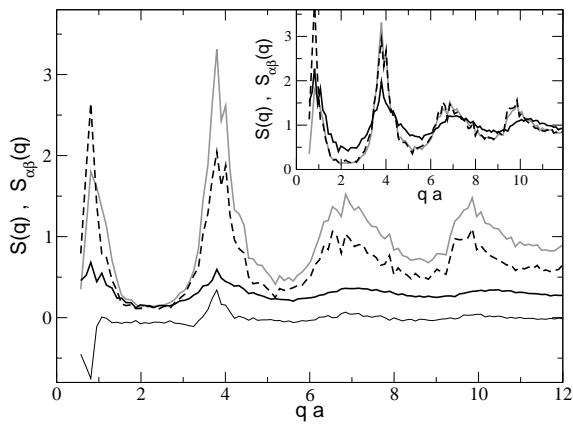}
\caption {\label{sq} Structure factor (thick grey line), and partial structure factors for fast-fast particles (thick black line), slow-slow particles (dashed black line) and fast-slow ones (thin black line), for $\phi_c=0.40$ and $\phi_p=0.42$. Inset: Structure factors rescaled to $S_{\alpha\beta}(q\rightarrow \infty)\rightarrow 1$.}
\end{figure}

The structure factor of the system is presented in Fig. \ref{sq} for the same system as Figs. \ref{gel} and \ref{slices}, as well as the contributions from pairs of fast-fast, slow-slow and fast-slow particles. The partial structure factors are defined by:

\begin{equation}
S_{\alpha\beta}\:=\:\frac{1}{N} \sum_{k=1}^{N_{\alpha}} \sum_{l=1}^{N_{\beta}}
\langle \exp \left( i {\bf q}\cdot ({\bf r_k}-{\bf r_l}) \right) \rangle,
\end{equation}

\noindent where the first sum runs over the $N_{\alpha}$ fast (slow) particles, and the second one over the  $N_{\beta}$ fast (slow) ones, and $N$ is the total number of particles. The low angle peak observed in the total structure factor and in the fast-fast and slow-slow contributions is caused by the voids in the network of particles. In contrast, the fast-slow contribution, shows a minimum in this region, caused by the clustering of fast and slow particles. 

In the inset to this figure, the fast-fast and slow-slow contributions are rescaled by a factor $N/N_f$ and $N/N_s$, the inverse fraction of fast and slow particles, respectively. Whereas the slow-slow contribution reproduces the neighbors peaks of the total $S(q)$, the contribution from fast particles shows less pronounced maxima, indicating that there are fewer bonds between fast particles than average.

Looking at the images in Fig. \ref{slices}, the origin of these two populations can be related to the position of every particle. The particles in the borders of the network are likely to break their bonds and diffuse to longer distances, while those particles inside the network, are surrounded by many particles and cannot escape. However, it must also be considered that we have a polydisperse system, where small particles not only can go through smaller holes but also form weaker bonds (see Fig. \ref{potential}). Both possibilities can be tested by correlating the displacement of every particle with its number of neighbors, and with its size. 

\begin{figure}
\psfig{file=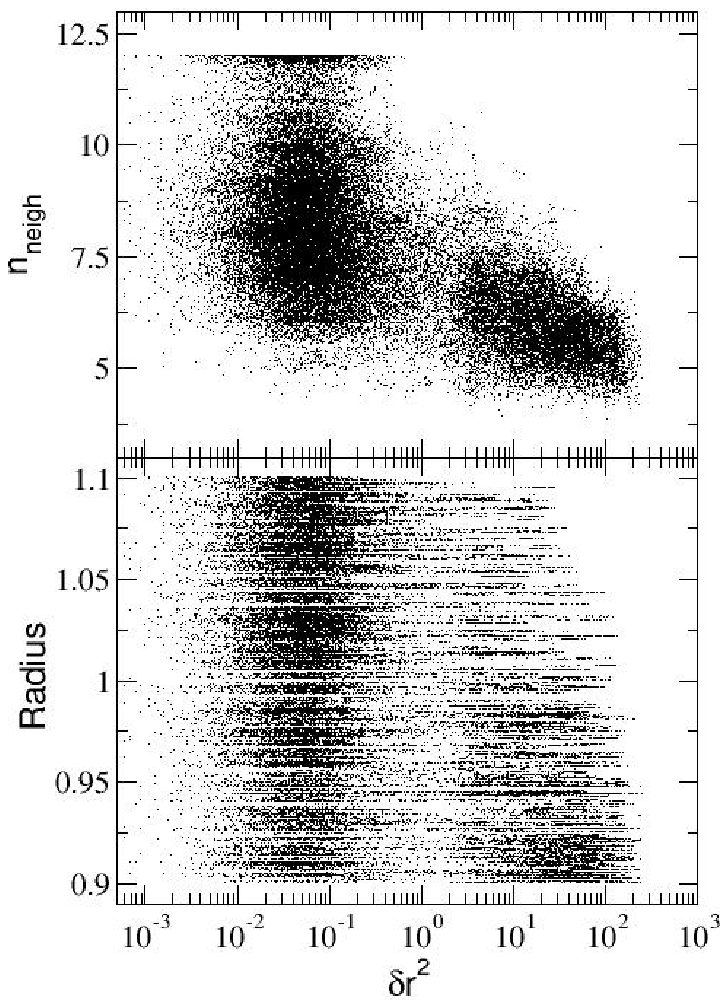,width=7.5cm}
\caption {\label{corr} Mean number of neighbors for every particle throughout the simulation (upper panel) and particle radius as a function of its squared displacement (lower panel) for $\phi_c=0.40$ and $\phi_p=0.42$.}
\end{figure}

Fig. \ref{corr} presents the mean number of neighbors of every particle averaged during the simulation up to time $t^*$ (upper panel) and its radius (lower panel) as a function of the squared displacement. Fast particles are mainly smaller and have fewer neighbors than the slow ones, as expected. However, it is also noticeable that there are also many small particles which are slow, indicating that discrimination by size has not completely occurred. On the other hand, many slow particles have as few bonds as the fast ones, showing that they are in the surface of the network but not breaking them and moving away. The mean number of neighbors of fast and slow particles are presented in Table II, where the distinction between fast and slow neighbors are made. It can be observed that fast particles have fewer neighbors than slow ones, and that the neighbors of slow particles are mainly slow ones, whereas fast particles are surrounded by more fast than slow particles.

\vspace{0.5cm}
\begin{center}
\begin{tabular}{cccc} \hline\hline
 & $\hspace{0.25cm}$ Total $\hspace{0.25cm}$ & $\hspace{0.25cm}$ Fast $\hspace{0.25cm}$ & $\hspace{0.25cm}$ Slow $\hspace{0.25cm}$ \\ \hline
$\hspace{0.25cm}$ Fast $\hspace{0.25cm}$ & 6.07 & 3.74 & 2.33 \\
Slow & 8.37 & 1.06 & 7.31 \\ \hline\hline
\end{tabular}
\end{center}
Table II: Total number of neighbors of fast and slow particles, and number of fast and slow neighbors.
\vspace{0.5cm}

\begin{figure}
\psfig{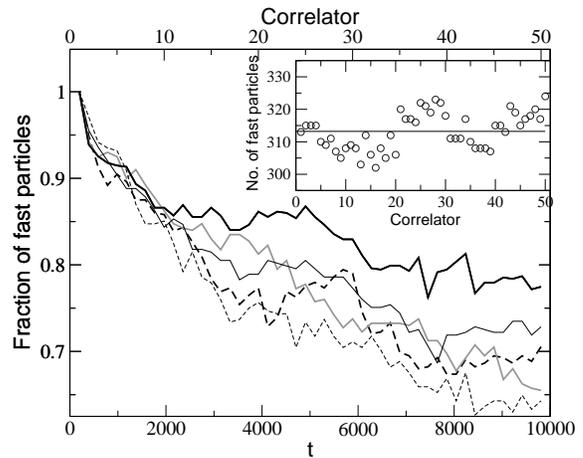}
\caption {\label{persistance} Fraction of fast particles in every correlator that were fast in the first correlator for $\phi_c=0.40$ and $\phi_p=0.40$ (thick black line), $\phi_p=0.41$ (thick black dashed line), $\phi_p=0.415$ (grey line), $\phi_p=0.42$ (thin solid line) and $\phi_p=0.425$ (thin dashed line) as a function of correlator (and time for $\phi_p=0.42$). Inset: Number of particles in every correlator for $\phi_c=0.40$ and $\phi_p=0.42$.}
\end{figure}

Because there is no intrinsic difference between fast and slow particles, and the existence of potential fast particles which are slow (with a only a few neighbors), it is possible that an exchange of particles between these two populations can be taking place. For this purpose, we have measured 50 correlators, each one started when the mean squared displacement of the previous one is equal to $a$, the mean radius. In every correlator the number of fast and slow particles can be determined by studying the distribution of squared displacements at time $t^*$, as shown in Fig. \ref{fig1}. In order to study the exchange between fast and slow particles, we have measured the fraction of fast particles which were fast in the first correlator. The results are presented in Fig. \ref{persistance} as a function of the correlator, for different polymer fractions. To make easier interpretation of the figure the time corresponding to the first correlator is also given for the case $\phi_p=0.42$, and the delay time between consecutive correlators, $\Delta t$ is provided in Table I.

Although the fraction of fast particles in every correlator that were fast in the first one decreases with time, the number of fast particles is constant throughout the simulation, as shown in the inset for $\phi_p=0.42$. This implies that indeed there is an exchange between both populations. However, after 50 correlators, only $30\%$ of the fast particles have formed enough bonds to become slow, which for the $\phi_p=0.42$ case amounts to about 100 particles. From the point of view of the ca. 700 slow particles, one hundred of them have been able to break their bonds and difuse away. 

From one correlator to the next, only about 15 particles change their character, while the mean squared displacement is equal to the mean particle radius. It should be noted, however, that the time between correlators, $\Delta t$ is much lower than $t^*$, the ratio between both times being approximately equal to 15 in all states. Thus, Fig. \ref{persistance} shows that the exchange between fast and slow particles in time $t^*$ is below $20\%$ of the amount of fast particles (about 60 particles for $\phi_p=0.42$). 

Since this exchange between fast and slow particles is so slow within every correlator, both species can be considered as being distinguishable and stable. Thus, we will analyze different dynamical properties of both populations as a function of time. In this analysis, the distinction between fast and slow particles is made at time $t^*$. Thus, from time $t=0$ to $t^*$, for every correlator, slow particles are always slow, although at longer times some of them may become fast.

\begin{figure}
\psfig{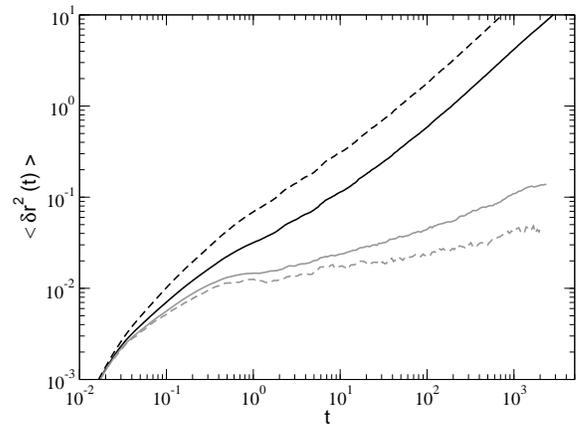}
\caption {\label{msd} Mean squared displacement of the system (solid black line), considering only the fast particles (dashed black line), slow particles (solid grey line) and the 15\% slowest particles (dashed grey line).}
\end{figure}

\subsection{Dynamics} 

The mean squared displacement of the fast and slow particles is presented in Fig. \ref{msd} for the state at $\phi_p=0.42$, as well as the average over the whole system. Fast particles are faster than slow ones at all times, even at extremely short times, when the displacement is still within the interaction range. The curve for slow particles shows that their motion is hindered at distances about $10^{-2}a$, and that a subdifusive motion takes over at long times. In the figure we have also included the mean squared displacement of the 15\% slowest particles, which is very similar to the curve of the slow particles, showing that the distribution of squared displacements of the slow particles is very narrow.

\begin{figure}
\psfig{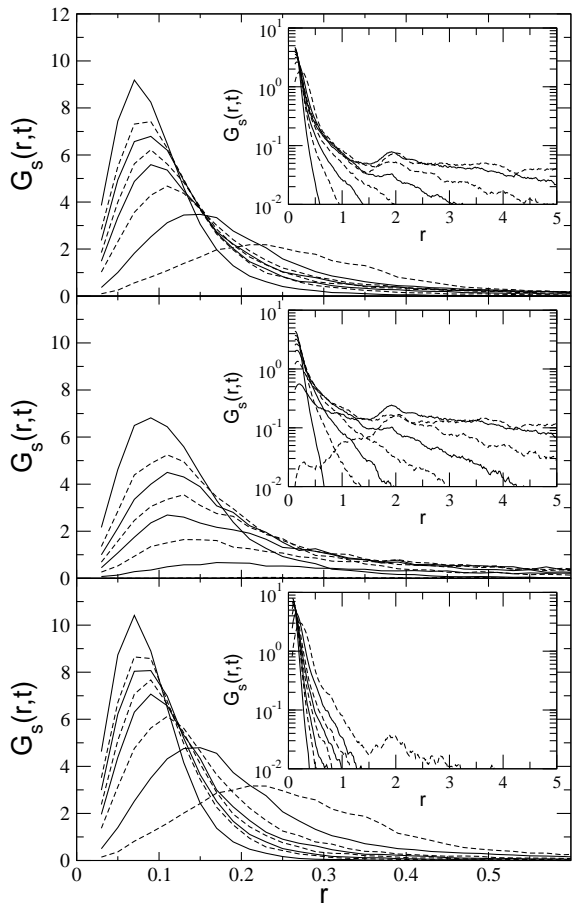}
\caption {\label{vanHoveself} Self part of the van Hove function for $\phi_c=0.40$ and $\phi_p=0.42$ (upper panel) for exponentially increasing times, $t_n=0.25\cdot 2^{2(n-1)}$, for the fast particles only (intermediate panel) and slow particles only (lower panel). Insets: The same as in the main graph in logarithmic scale.}
\end{figure}

To gain more information on the motion of particles, we have analyzed the self part of the van Hove function, defined as:

\begin{equation}
G_s^{\alpha}(r,t)=\frac{1}{N_{\alpha}} \left\langle \sum_{i=1}^{N_{\alpha}} \delta \left(r-\left| {\bf r_i}(t)-{\bf r_i}(0)\right| \right) \right\rangle
\end{equation}

\noindent where, again, $N_{\alpha}$ is the number of particles of type $\alpha$ in the system. This function provides information about the motion of single particles, and its deviation from Gaussian behavior is quantified by the non-Gaussian parameter, $\alpha_2$ \cite{rahman64}. The self part of the van Hove function for the system at $\phi_c=0.40$ and $\phi_p=0.42$ is presented in Fig. \ref{vanHoveself} (upper panel) for exponentially increasing times. 

The van Hove function for the system shows the typical behavior of close-to-glass systems; at short and intermediate times, the van Hove function has a maximum at the estimated localization length, of the order of the cage size, or the interaction range in our case. At long times, the particles break free of their cages (network of bonds) and diffuse to longer distances. However, when plotted in logarithmic scale, see the inset to upper panel, it can be observed that a non-negligible tail appears at long distances. This feature is caused by particles moving at long distances even at short times, i.e. the fast particles. Thus, the van Hove function has been calculated considering fast particles only (intermediate panel), and slow particles only (lower panel). It is worth remembering that the distinction between fast and slow particles was made when the mean squared displacement was equal to $10 a^2$, which for this state corresponds to $t^*=2871$. 

Indeed, the fast particles are responsible for the tail at long distances, but also fewer fast particles are trapped in the network of bonds (see the height of the peak) and those which are imprisoned escape earlier than average. On the other hand, there are more slow particles confined, and for much longer times, and their van Hove function does not present the long distance tail. It is interesting to observe that at long times, there are just a few fast particles still close to their original positions and the van Hove function is practically flat (note the dashed line in the second panel close to the x-axis), whereas most of the slow particles are still at a distance of the order of the interaction range. The latest time presented, $t=4096$, is after the distinction between fast and slow particles was made, and therefore, some exchange between the fast and slow populations may have taken place. This is the origin of the tail at long distances observed in the function for the slow particles at this time. 

\begin{figure}
\psfig{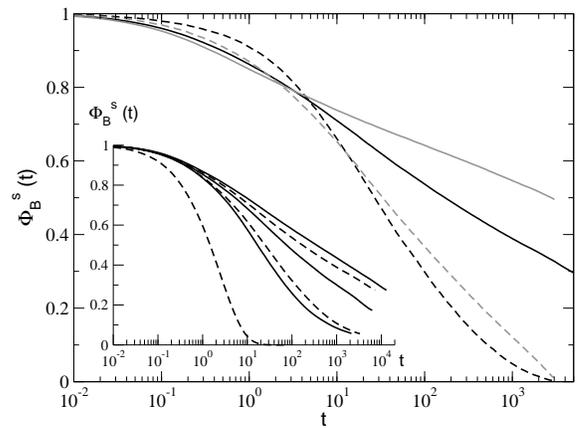}
\caption {\label{phiB} Bond correlation function for $\phi_c=0.40$ and $\phi_p=0.42$ for the whole system (solid black line), for bonds between fast particles (dashed black line), bonds between slow particles (solid grey line) and bonds between fast and slow particles (dashed grey line). Inset: Bond correlation function for the systems at $\phi_c=0.40$ and $\phi_p=0.30, 0.40, 0.41, 0.415, 0.42, 0.425$ (from left to right).}
\end{figure}

Since the driving process for the gel transition is bond formation, we study now the differences between fast and slow particles from the point of view of their bonds and of their environment. The bond correlation function, $\Phi_B(t)$, is defined as the fraction of bonds that have uninterruptedly existed from time $t=0$. This function is presented in Fig. \ref{phiB} for different states approaching the gel transition (inset to the figure). As the polymer fraction increases, the bonds between particles are more stable, not only because of the increasing attraction strength, linear with $\phi_p$, but also due to the higher number of neighbors, forming an intricate network. This collective character becomes apparent also since this function departs from the simple exponential more and more.

In the main body of Fig. \ref{phiB} the bond correlation function for $\phi_p=0.42$ is analyzed in more detail, presenting the contributions from bonds between two fast particles, two slow ones and bonds between fast and slow particles. The fast-fast bonds break faster at intermediate and long times, showing that these particles move as single particles, and not in clusters of fast particles. In contrast, the most stable bonds are those formed between two slow particles, since they hardly move. However, this trend is inverted at short times; there, the bonds between slow particles are the easiest to break, while those between fast ones are more stable. This feature can be rationalized considering that slow particles live in a rigid environment, where the motion of a single particle, even if small, can break bonds, as the others are not able to reaccommodate to keep their bonds alive. For fast particles, the loose environment allows some fluctuation, and their bonds are more stable at these short times.

\begin{figure}
\psfig{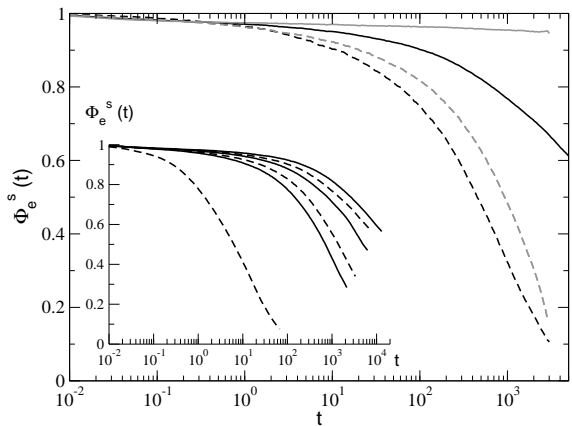}
\caption {\label{phie} Environment correlation function for $\phi_c=0.40$ and $\phi_p=0.42$. The lines have the same meaning as Fig. \ref{phiB}. Inset: Environment correlation function for the states as Fig. \ref{phiB}.}
\end{figure}

Although the bonds between slow particles are the most fragile at short times, the environment of these particles does not change, i.e. their likely neighbors are the same for very long periods of time. This fact can be recognized by studying the fraction of neighbors of the particle that were neighbors at $t=0$. This environment correlation function, $\Phi_e(t)$ is plotted in Fig. \ref{phie} for different states (inset to the figure) and in more detail for the $\phi_p=0.42$ case. Because this function measures the relaxation of the environment of every particle, instead of the lifetime of the bonds, it decays much slower than $\Phi_B$, as observed in comparing Figs. \ref{phiB} and \ref{phie}.

As expected, the environment correlation function decays slower as the attraction strengthens, and in the slowest case, $\phi_p=0.425$, it has decayed to about $\Phi_e=0.5$ only, although the mean squared displacement was $\langle \delta r^2 \rangle=20 a^2$. The contribution to the global behavior from fast neighbors of fast particles decays much faster than average, as that of slow-fast pairs (or vice versa). In contrast, the neighborhood of the slow particles formed by other slow particles is hardly altered, keeping more than $95\%$ of their neighbours in average. It is important to note that although the bonds between slow particles can break, their neighbours do not change. The system, thus, is constantly rearranging, breaking and forming new bonds, but the overall structure is quite stable. 

The behavior of the correlation functions $\Phi_B(t)$ and $\Phi_e(t)$ for the bonds and environment of the fast particles, indicates that these particles break their bonds with their neighbors and change them, showing that their fast motion does not take place inside fast clusters. Nevertheless, since the process of breaking single bonds is much faster than neighborhood relaxation, the restructuration of bonds alluded above also takes place for these particles. In order to understand in more detail the behavior of these particles, we have analyzed the distinct part of the van Hove correlation function, defined as:

\begin{equation}
G_d^{\alpha \beta}(r,t)=\frac{N}{N_{\alpha}N_{\beta}} \left\langle \sum_{i=1}^{N_{\alpha}} \sum_{j=1}^{N_{\beta}} \delta \left( r- \left| {\bf r_i}(t)-{\bf r_j}(0)\right| \right) \right\rangle
\end{equation}

\noindent where $N, N_{\alpha}$ and $N_{\beta}$ have the same meaning as above. For $\alpha=\beta$, the fraction in front of the summation reads $N/N_{\alpha}(N_{\alpha}-1)$, and the case $i=j$ is omitted in the summation. For long times, this function tends to one, $G(t\rightarrow \infty) \rightarrow 1$, as the structure of the system at $t=0$ relaxes.

\begin{widetext}
\begin{center}
\begin{figure}
\psfig{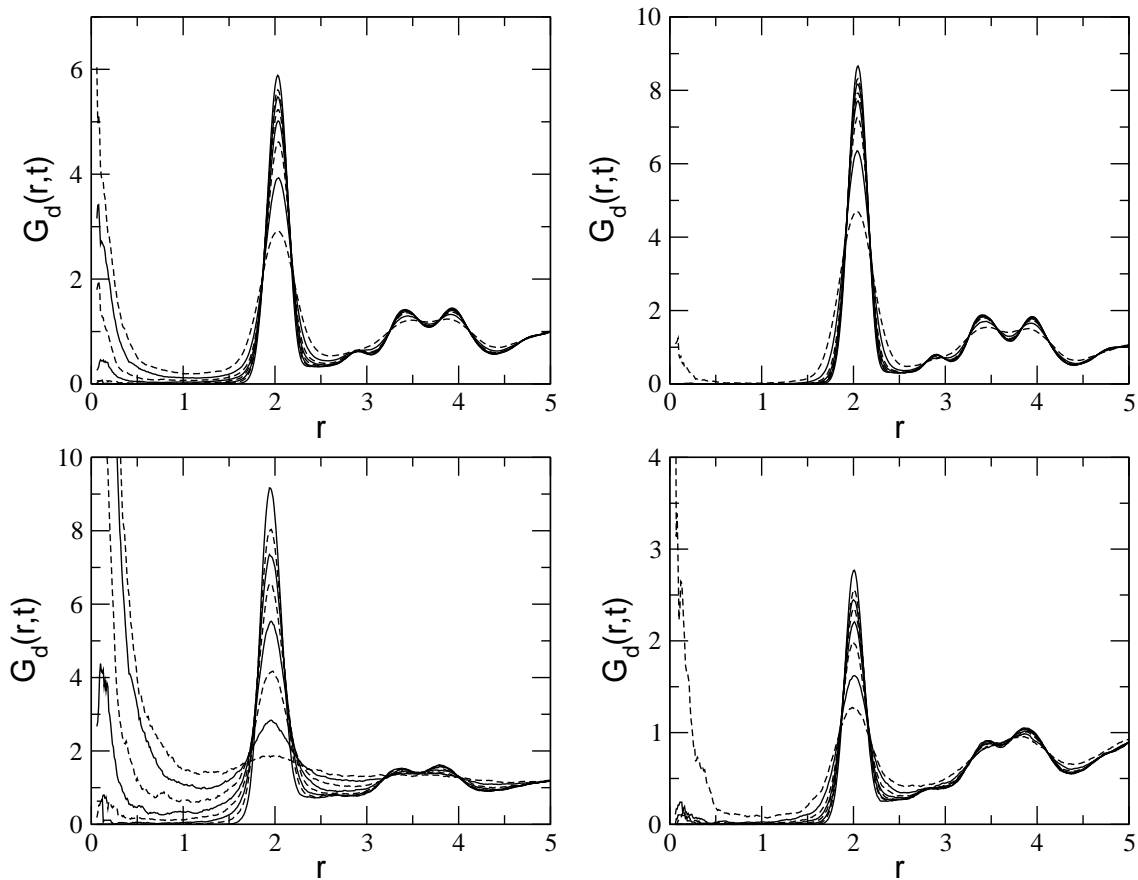}
\caption {\label{vanHovedist} Distinct part of the van Hove function for $\phi_c=0.40$ and $\phi_p=0.42$ (upper-left panel) for exponentially increasing times, $t_n=0.25\cdot 2^{2(n-1)}$, for the slow-slow particles only (upper-right panel), fast-fast particles (lower-left panel) and slow-fast pairs (lower-right panel).}
\end{figure}
\end{center}
\end{widetext}

In Fig. \ref{vanHovedist} we present this function for the same times as Fig. \ref{vanHoveself} for the whole system as well as the contributions from the different combinations. As expected from the discussion above, the function for the whole system does not reach its limit at long times, since the structure does not relax in the time of the simulation. It is also interesting to note that the function shows a peak at the origin, whose height grows with time. This peak is caused by particles occupying exactly the same place where other particles were at previous times. In glasses, this feature is recognized as a signature of hopping, or string-like motions \cite{donati99b}.

The origin of the peak at $r=0$ can be understood studying the contributions from the different pairs of particles. The peak appears only in the fast-fast contributions, though it is present at times longer than $t^*$ in the other contributions. It can be rationalized as fast particles occupying the place where other fast particles laid, but string-like motions are not observed (by direct visual observation of the simulation). The mean number of neighbors of these particles while occupying the former place of another fast particles is $6.54$, with $3.07$ fast neighbors and $3.48$ slow ones. The ratio of slow-fast neighbors of these particles is much larger than the average over all of the fast particles (see table I), showing that they are much closer to (in contact with) the structure formed by the slow particles. We thus see preferential sites in the boundary of this network, where fast particles can attach. However, at long times, the function shows the expected limit, $G(t\rightarrow \infty) \rightarrow 1$, except for the peak at the origin, showing that the structure formed by the fast particles has disappeared.

The distinct part of the van Hove function from slow-slow pairs of particles shows little evolution with time, as expected, since the structure formed by these particles is quite rigid and stable. For $t=4096$, $t>t^*$, a low peak at the origin appears, indicating that hopping can be detected even for the slow particles for long times. The slow particles that change to fast ones, or vice versa, cause the peak at $r=0$ in the contribution from the fast-slow pairs. The neighbor peak is in this case lower than in the other contributions, since there are few contacts between fast and slow particles. 

\begin{figure}
\psfig{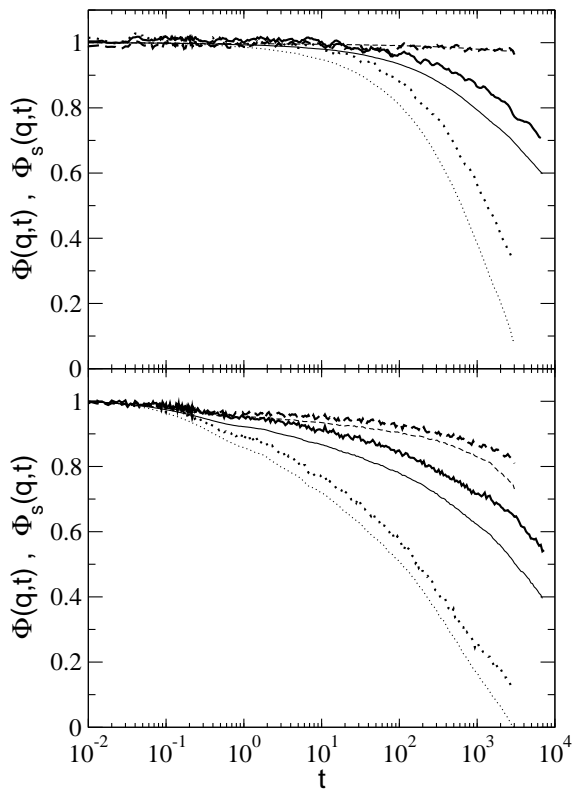}
\caption {\label{phi} Coherent (thick lines) and incoherent (thin lines) density correlation function for $\phi_c=0.40$ and $\phi_p=0.42$, at $qa=1$ (upper panel) and $qa=3.9$ (lower panel). Contribution from the slow particles (dashed lines) and from the fast ones (dotted lines) are presented.}
\end{figure}

Finally, we will analyze the relaxation of the structures formed by the fast and slow particles. Figs. \ref{phiB}, \ref{phie} and \ref{vanHovedist} have already provided information about this feature, though studying it locally. The relaxation of a structure is better observed by the (coherent) density-density correlation function:

\begin{equation}
\Phi^{\alpha \beta}(q,t)=\frac{1}{N}\left\langle \sum_{i=1}^{N_{\alpha}} \sum_{j=1}^{N_{\beta}} \exp \left\{ i {\bf q}\cdot \left[ {\bf r_i}(t)-{\bf r_j}(0) \right] \right\} \right\rangle
\end{equation}

\noindent where the brackets indicate ensemble averaging, and ${\bf q}$ is the wave vector, giving the typical size of the structure. The self or incoherent part of this function, $\Phi_s^{\alpha}(t)$ is calculated taking $i=j$. Both the coherent and incoherent correlation functions are plotted in Fig. \ref{phi} for the system at $\phi_c=0.40$ and $\phi_p=0.42$ at the wavevectors of the first peak in the structure factor, $qa=1$, (upper panel) and at the neighbor peak, $qa=3.9$ (lower panel). The former value has been taken as representative of the macro-structure of the colloidal gel, whereas the latter gives again information about the local environment of the particles.

The structure formed by the slow particles indeed changes very little in the time of the simulation, as observed in their contribution to the coherent part of the intermediate scattering function at $qa=1$, which decays less than a $5\%$ of its initial value. In contrast, the contribution from fast particles, decays much faster, as expected from the results above. It is interesting to compare the coherent and the incoherent parts of the function. At this wave vector, the self parts decay a decade faster than the coherent parts. For the higher wave vector, $qa=3.9$, the correlation functions decay faster than at the previous $q$, implying that the structures at the corresponding length relax faster than the overall colloidal structure. Again, the contributions from slow and fast particles show the expected behaviour, although the relaxations are faster in this case. 

\subsection{Discussion}

The results presented in the previous subsections lead to a dynamically heterogeneous picture of a colloidal gel. In a simple description, it can be considered as composed by two distinct populations of particles: 'fast' and 'slow'. Although there is no intrinsic difference between both particles, the interchange between them has a very low rate. The slow particles form a  rigid network of particles, very stable with time; the others, quite movable, move using sites where other fast particles laid, since the structure formed by these particles relaxes ten times slower than the mean time for particle relaxation. As the attraction strengthens, i.e. the gel transition is approached, the distinction between fast and slow particles becomes more pronounced, and the fraction of fast particles decreases, as observed at fixed mean squared displacement. 

These results were tested by analyzing the evolution of the squared displacement distribution up to longer times ($10^4$ time units). Only a few correlators were considered. For the system at $\phi_p=0.42$, the peak of fast particles moves continuously to longer distances reaching the maximum distance in the simulation (one half of the box edge $ \sim 120 a^2$). The peak of the slow particles slowly moves to higher distances, but stays of the order of $10^{-1}a^2$. This distinction up to long times indicates that the structure of the slow particles does not relax in the simulation time-window, as observed in the intermediate scattering function, Fig. \ref{phi}. 

This fact may raise the point of how well equilibrated our systems are. The slowest system, $\phi_p=0.42$, was equilibrated for $5\cdot 10^4$ time units, which is more than ten times $t^*$ for this state. The correlation functions measured at different times, the ratio of fast/slow particles, as well as other quantities, did not show any dependency on the initial time (between $5\cdot 10^{4}$ and $6\cdot 10^{4}$). Apparently, this state is, thus, well equilibrated. For lower polymer fractions, $\phi_p \leq 0.41$, the whole system relaxes within the simulation time-window. We cannot ascertain, but expect that the coherent and incoherent correlation functions of Fig. \ref{phi} decay via the exchange of fast and slow particles shown in Fig. \ref{persistance}. If so, these correlation functions should decay eventually to zero, so that the systems we have studied remain on the liquid rather than the gel side of the nonergodicity transition, as assumed in the analysis we have made in this paper, and previously \cite{puertas03}. However, since the contributions from slow particles to the correlation functions decay only a little within the observable time window, we cannot rule out a scenario where the correlation functions decay not to zero, but to a finite value, at timescales beyond those studied here. Such a plateau could arise in the presence of a persistent structural component comprised of a subset of the slow particles. There is no way to rule out such a scenario other than by extending the simulation runs by at least one further decade in time, which is beyond the scope of this paper.

It is interesting to note, nevertheless, that the overall behavior of the system is correctly described by MCT \cite{puertas03}. The localization length obtained in the simulations and in the theory are in agreement, showing that bond formation is the leading process in colloidal gelation. The behavior of the self part of the density correlation functions obey the results obtained in MCT, both in time-scaling, and the specific form of the decay. The behaviors of the time scale of the correlation function as a function of the attraction strength and of the wavevector, also fit in the predictions from MCT. However, MCT gives information through correlation functions \cite{gotze91,gotze92,hansen86}, and thus cannot give insights into the origins of the dynamical heterogeneities presented here. 

The fast particles fluidize the system, and allow for restructuring of parts of the system, causing a decay in the correlation function, which can be described using MCT. As the gel transition is approached, there are more and more slow particles, forming the rigid network, but from our observations, we cannot say what are the limits of the fractions of fast or slow particles at gelation. Whereas MCT is able to describe the behavior of the system close to the gel transition, though only for averaged quantities, it is expected that the dynamical heterogeneities described here will become important closer to the transition, and deviations from the MCT behavior should be expected.

The strong dynamical heterogeneities presented here appear to be caused by the structural heterogeneities in the system. Because the latter are absent in the glass transition driven by steric hindrance, we do not expect such big effects in this transition. The string-like motions reported in, e.g. Lenard Jones systems, which also fluidize the system, are caused also by structural heterogeneities, although of the size of a single particle. Thus, their effects in the behavior of the system is not as dramatic as in the gel case. Presumably, the effects of the two populations will be more important for gelation at lower packing fractions, where the structural heterogeneities are also more important.

\section{Conclusions}

We have studied in this paper the dynamics of a colloidal gel, showing that its microscopic behavior is far from homogeneous. Some particles are faster than others, and two populations can be discerned, according to their squared displacement at long times. Fast particles have fewer neighbors than slow ones, and thus, can escape their bonds easier. Because there is no structural difference between them, some fast particles become slow, when trapped by others, and some slow ones finally break free, and become fast. However, the exchange rate between fast and slow particles is very low and allows analyzing the system as composed of two different populations. Indeed, this oversimplified two population picture gives an adequate description of the main features of the system.

The motion of the fast particles does not occur in clusters or arms rearranging, but they move singly, as observed in the bond correlation function. Their bonds are the weakest at intermediate and long times, and the structures formed by the fast particles relax faster than average. In contrast, in the network formed by the slow particles the bonds between slow particles are very fragile at short particles due to the stiffness of the network. The intermediate scattering function at the wave vector related to the structure of the gel, shows that this structure is very stable with time, while the structure formed by the fast particles relaxes much quicker. 

These observations on the microscopic dynamics of the gel cannot be accounted for in MCT in its present form, although the average behavior is correctly described by this theory. Close to the gel transition, we expect, the effects caused by the distinct populations would dominate the behavior of the system and the MCT picture would break down.

\begin{center}
{\sc Acknowledgments}
\end{center}

The authors thank F. Sciortino for useful and stimulating discussions. A.M.P. acknowledges the financial support by the CICYT (project MAT2003-03051-CO3-01).

\end{document}